\documentstyle[prl,aps,multicol,epsf]{revtex}

\tighten
\begin{document}

\draft
\title{Interaction-induced oscillations of the tunneling density of states
  \\in a non-quantizing magnetic field}
\author{A.M. Rudin, I.L. Aleiner$^*$, and L.I. Glazman}
\address{Theoretical Physics Institute, University of Minnesota, Minneapolis MN
55455}
\maketitle

\begin{abstract}
We study tunneling into interacting disordered two-dimensional electron
gas in a non-quantizing magnetic field, which does not cause the standard
de Haas
-- van Alphen oscillations.
Interaction induces a new type of oscillations in the tunneling density of
states with the
characteristic period  of cyclotron quantum  $\hbar\omega_c$.
\end{abstract}

\begin{multicols}{2}

It is well-known that  strong magnetic field modifies the single-particle
density of
states (DOS) of non-interacting electrons due to the Landau quantization.
In a two-dimensional
electron  gas the quantization leads to a peak structure in the DOS, which
is  revealed in tunneling
experiments as peaks in the dependence of the tunneling conductance  on  applied
bias, see, {\it e.g.}, Ref.
\cite{Eisenstein}. The form and the width of these peaks are
determined\cite{Ando} by the disorder.
Experiments\cite{Eisenstein,Ashoori,Turner} show also suppression of the
conductance at zero bias.
This suppression  is a manifestation of interaction between electrons and
has been studied
theoretically both for disordered
\cite{Girvin} and for clean\cite{People} two-dimensional conductors in the
quantizing magnetic
field.

In a weak  magnetic field  the distance between the Landau levels,
$\hbar\omega_c$, is smaller
than  their disorder-induced width. In such ``classical'' magnetic field
the  Landau quantization peaks in DOS disappear, while  the aforementioned
interaction correction
to tunneling DOS survives and in the limit of zero field evolves into a
singular at the Fermi energy
negative logarithmical correction  predicted originally by  Altshuler,
Aronov, and Lee
\cite{Altshuler}.

The question arises, whether or not in such  ``classical'', $\omega_c
\tau_s \ll 1$, limit the
magnetic field  influences the spatially averaged density of electron states
measured in tunneling experiments (here $\tau_s$ is
the  electron quantum lifetime). For the non-interacting
system, effect of the weak magnetic field on DOS is exponentially small
\cite{Ando}, $\propto
\exp[-2\pi/(\omega_c
\tau_s)]$,  and can be neglected. The goal of the
present paper is to show that, to the contrary to the non-interacting case,
in the {\it interacting}
electron gas the ``classical''  magnetic field does produce a significant
effect on tunneling DOS.
This effect becomes  pronounced if  the disorder potential is weak enough and
smooth, with the correlation length much larger than the Fermi wave length and
the amplitude much smaller than the Fermi energy. In such a potential,
electrons experience small-angle scattering, and their transport relaxation
time
$\tau_{\rm tr}$, is much larger than $\tau_s$. Thus
there exists  range of magnetic fields, in which Landau
quantization is suppressed ($\omega_c \tau_s \ll 1$), while classical
electron trajectories are strongly affected by the field ($\omega_c \tau_{\rm tr}\gg
1$). In this regime interaction
correction to the tunneling DOS, $\delta \nu (\epsilon)$, is strongly
enhanced with respect to the zero magnetic field case. Furthermore, we will
show that it
 exhibits peaks as a function of
energy with the distance between peaks equal to the
cyclotron quantum, $\hbar\omega_c$.  The further a peak is away from the
Fermi level, the smaller
and wider it is. The shape
of the $n$-th peak in DOS, $|\epsilon - n\hbar \omega_c| \lesssim \hbar
\omega_c/2$, is given by:
\begin{equation}\label{result}
{\delta \nu (\epsilon) \over \nu} = {\hbar(\omega_c \tau_{\rm tr})^2\over
8\pi\epsilon_F \tau_{\rm tr}} \,  {1 \over n} \,
f \!\left({\epsilon -n \hbar \omega_c \over \hbar  n^2 /\tau_{\rm tr}} \right),
\end{equation}
where
\begin{equation}\label{result1}
f(x) = {1\over \sqrt{2}} \left[{1+\sqrt{x^2+1}\over x^2+1} \right]^{1/2},
 \end{equation}
energy $\epsilon$ is measured from the Fermi level, and $\nu = m/\pi\hbar^2$
is the free-electron density of states ($m$ is the electron mass).  The
peaks overlap strongly for
$\epsilon \gtrsim \hbar\omega_c\sqrt{\omega_c\tau_{\rm tr}}$, and the
oscillatory structure is washed
out.

Sensitivity of tunneling DOS to the classical magnetic field comes from the
fact, that, as we
will show,  the interaction  correction to tunneling DOS is associated with
the self-crossing of
classical  electron trajectories. We denote the probability for an electron
to complete a loop of
self-crossing trajectory over time $t$ as $K(t)$. The interaction
correction to DOS, $\delta \nu
(\epsilon)$, turns out to be related to the Fourier transform of this
probability, $\delta \nu
(\epsilon) \propto K(\epsilon) = \int_0^{\infty} dt e^{-i\epsilon t} K(t)$.
The  strong enough,
$\omega_c\tau_{\rm tr}
\gg 1$, magnetic field curves the electron trajectories,  significantly
affects the return
probability and, in turn, affects the tunneling DOS.

For long time scales $t \gg  \tau_{\rm tr}$, the  function $K(t)$ can be found
from the diffusion
equation. It gives $K(t)
\propto (Dt)^{-1}$ for the two-dimensional case ($D$ is the diffusion
coefficient).  The Fourier transform, $K(\epsilon)$, is proportional to  $
\ln(\epsilon)$, which leads  to a  predicted by
Altshuler, Aronov and Lee \cite{Altshuler} logarithmic correction to DOS at
small energies, $\epsilon
\ll \hbar /\tau_{\rm tr}$, with the renormalized by the magnetic field
diffusion coefficient
\cite{Levitov}.

At short time scales, $t\ll
\tau_{\rm tr}$, electrons move ballistically along the cyclotron orbits.
Provided that
$\omega_c\tau_{\rm tr} \gg 1$, during the time
$t$ electron may return to the initial point many times. Thus at these
short time scales the
magnetic field significantly increases the return probability $K(t)$. Multiple
periodic returns of electron  produce  peaks in the probability Fourier
transform
$K(\epsilon)$ at energies, which are multiples of the cyclotron quantum.
Tunneling
density of states oscillates with the same period, which is reflected by
Eq. (\ref{result}).

Now we derive expression for the interaction correction to DOS valid for
arbitrary energies. We will put
$\hbar=1$ in all intermediate formulas. In an ideal pure sample the
electron density
does not depend on coordinates,
$n_e({\bf r}) = n_0$.   Due to the sharp Fermi edge, scattering of the
electron forming the Fermi sea
on impurities results in an interference pattern in electron density. This
pattern is commonly referred
to as  the Friedel
oscillation\cite{Mahan}. In general, one can express  density profile of
noninteracting electrons in
terms of  the exact retarded Green function,  ${\cal
G}^R_\epsilon ({\bf r}, {\bf r}') = \sum_l \psi_l^* ({\bf r}')\psi_l ({\bf
r})/(\epsilon -
\epsilon_l +i0)$ , of an electron in the random potential:
\begin{equation}\label{n(r)}
n_e({\bf r}) = -{2\over\pi}\,\int_{-\epsilon_F}^0 d\epsilon \,  {\rm
Im}{\cal G}^R_\epsilon ({\bf r}, {\bf r}).
\end{equation}
 Single electron wave function $\psi_l({\bf r})$ satisfies the
Schr\"odinger equation  for noninteracting electrons, $\hat H_0 \psi_l =
(\epsilon_l+E_F)\psi_l$, where $\hat H_0 = -(\hbar^2 / 2 m ) \nabla^2 + U_{
r}({\bf
r})$, $E_F$ is the Fermi energy, and $U_{r}({\bf r})$ is the random potential.
 In the presence of interaction $V({\bf r}-{\bf
r'})$ between electrons, the Friedel oscillation produces an additional
term in the Hamiltonian, $\hat H_{HF}$, which can be
presented (see {\it e.g.} Ref. \cite{Mahan}) as a sum of  Hartree, $V_H$,
and exchange, $V_F$,  terms:
\begin{eqnarray}
 H_{HF} ({\bf r}, {\bf r}') &=& V_H ({\bf r}) \, \delta ({\bf r}-{\bf r}')
+ V_{F} ({\bf r}, {\bf r}') \\ \label{Hartree}
V_H ({\bf r}) &=& \int V({\bf r} -{\bf  r}'') \delta n_e({\bf r}'')d{\bf r}'' \\
\label{exc}
V_{F} ({\bf r}, {\bf r}') &=& {1 \over 2} \,  V({\bf r} - {\bf
r}') \delta\rho ({\bf r},{\bf r}').
\end{eqnarray}
Here $\delta\rho({\bf r}, {\bf r}')$ is the perturbation of the density
matrix, $\rho({\bf r},
{\bf r}') =  -(2/\pi)\,\int d\epsilon \,  {\rm Im}{\cal G}^R_\epsilon ({\bf
r}, {\bf r}')$,
by the random potential.  Only  the electrons with the same spin
participate in the exchange
interaction, which is reflected by the factor $1/2$ in Eq.~(\ref{exc}).  The
Hartree-Fock energy (\ref{Hartree})-(\ref{exc}) oscillates as a function of
coordinate  in the
same manner as $\delta n_e({\bf r})$ does.

The local DOS is related to the retarded Green
function of electron,  $\nu (\epsilon, {\bf r}) = -(2/\pi)\,  {\rm Im}
{\cal G}^R_{\epsilon} ({\bf r}, {\bf r})$.  Scattering of electron on
the Hartree-Fock potential, Eqs.~(\ref{Hartree})-(\ref{exc}),  induces a
correction to  ${\cal G}^R_{\epsilon} ({\bf r},  {\bf r})$,  which, in the
Born approximation, can
be expressed as:
\begin{eqnarray}\label{deltaG}
\delta {\cal G}^R_{\epsilon} ({\bf r}, {\bf r}) &= &
\int {\cal G}^R_{\epsilon}({\bf r}, {\bf r}')  V_H({\bf r}')
 {\cal G}^R_{\epsilon}({\bf r}', {\bf r})  d{\bf r}' \nonumber \\
& -& \int {\cal G}^R_{\epsilon}({\bf r}, {\bf r}')  V_{F}({\bf r}',{\bf r}'')
 {\cal G}^R_{\epsilon}({\bf r}'', {\bf r})  d{\bf r}' d{\bf r}''.
\end{eqnarray}
We will be interested in the spatially averaged density of states, $\delta
\nu (\epsilon) =
(1/{\cal S})\int \delta \nu (\epsilon, {\bf r})d{\bf r}$,  where ${\cal
S}$ is the area of the system.   For simplicity we will start with the
case of the finite-range interaction potential and will calculate the
Hartree contribution to the averaged DOS. Making use of Eqs.
(\ref{n(r)}), (\ref{Hartree}), and (\ref{deltaG}),  and exploiting the
identity,
$
\int {\cal G}^R_{\epsilon}({\bf r}', {\bf r}) \, {\cal G}^R_{\epsilon}({\bf
r}, {\bf r}'')   d{\bf r} = \partial {\cal G}^R_{\epsilon}({\bf r}', {\bf r}'')/
\partial
\epsilon,
$
we obtain:
\begin{eqnarray}\label{deltaNu}
\delta \nu_{H} (\epsilon) &=& {2\over \pi^2 {\cal S} }  {\rm
Re}\int_{-\epsilon_F}^0  d\epsilon_1  \int d{\bf r}' d{\bf r}'' V({\bf r}'-{\bf
r}'')
  \nonumber \\
& \times & {\partial   {\cal G}^R_{\epsilon}({\bf  r}',  {\bf r}')
\over \partial   \epsilon} [{\cal
G}^R_{\epsilon_1} ({\bf r}'',{\bf r}'' ) - {\cal G}^A_{\epsilon_1} ({\bf
r}'',{\bf r}'' )],
\end{eqnarray}
where ${\cal G}^A_{\epsilon} ({\bf r},{\bf r}' ) = [{\cal G}^R_{\epsilon}
({\bf r}',{\bf r} )]^*$.
We are interested in the correction to the density of states averaged
over the realizations of the disorder potential.  Average of the  product
of two retarded Green functions ${\cal G}^R_{\epsilon} ({\bf  r}',  {\bf
r}') {\cal G}^R_{\epsilon_1} ({\bf r}'',{\bf r}'' )$ does not contain
contributions associated with the  electron trajectories longer than
$\lambda_F$. Thus, this product does not  produce DOS  energy dependence
at small, as compared to $E_F$, energies, and can be neglected. On the contrary,
 averaged product  ${\cal G}^R{\cal G}^A$ is determined by  long
electron trajectories. Furthermore,  it can be expressed \cite{AL} in terms
of the classical
probability density ${\cal D}$:
\begin{eqnarray} \label{K(e,p)}
 \langle && {\cal G}^R_{\epsilon_1}({\bf r}_1,{\bf r}_2){\cal
G}^A_{\epsilon_2}({\bf
r}_3,{\bf r}_4)
\rangle = \pi\nu\int {d\phi_1 \over 2\pi}
\int {d\phi_2 \over 2\pi}  \nonumber \\
& & \times   e^{i{\bf p}_1({\bf r}_1 - {\bf r}_4)} e^{i{\bf p}_2({\bf r}_3
- {\bf r}_2)}
{\cal D}(\epsilon_1-\epsilon_2 ; {\bf
r}_1, \phi_1; {\bf r}_2, \phi_2);  \nonumber \\
& & {\cal D}(\omega ; 1; 2 ) = \int_0^\infty dt e^{i\omega t }{\cal D}(t;
1; 2 ).
\end{eqnarray}
Here ${\bf p}_i = p_F(\cos \phi_i, \sin\phi_i )$, and ${\cal D}(t;1;2)$ is
the probability
density for electron which starts at moment $t=0$ in point ${\bf r}_1$ with the
direction of momentum
$\phi_1$ to arrive at moment $t$ to the point ${\bf r}_2$ with momentum
direction
$\phi_2$.

Eq.~(\ref{K(e,p)}) is valid as long as the sizes $|{\bf r}_1-{\bf r}_4|$ and
$|{\bf r}_2-{\bf r}_3|$ of spatial domains defining the ends of a trajectory
are small enough, so that electron propagation in these two domains can be
described by plane waves.  Thus formula (\ref{K(e,p)}) is
valid if $\epsilon_F\tau_{\rm tr} \gg 1$ (semiclassical regime), and for the
arguments sufficiently close to each other pairwise; in general, the condition
$|{\bf r}_1-{\bf r}_4|, |{\bf r}_2-{\bf r}_3| \ll v_F\tau_{s}$ must be
satisfied, however, in the special case of Eq.~(\ref{deltaNu}), we need the
product of Green functions with ${\bf r}_1 = {\bf r}_3$ and ${\bf r}_2 = {\bf
r}_4$, and this condition can be eased. Indeed, the integral over the angular
variables in Eq.~(\ref{K(e,p)}) is dominated by the close to each other momenta
${\bf p}_1$ and ${\bf p}_2$, and therefore a weaker requirement, $|{\bf
r}_1-{\bf r}_4|, |{\bf r}_2-{\bf r}_3| \ll l_{\rm tr}, R_c$, should be satisfied. Here $R_c
= v_F/\omega_c$ is the cyclotron radius, $l_{\rm tr} = v_F\tau_{\rm tr}$ is the electron
transport relaxation length, and $v_F$ and $p_F$ are the Fermi velocity and
momentum respectively.  Further derivation of the  Hartree correction requires
substitution of Eq.~(\ref{K(e,p)}), with arguments ${\bf r}_1   = {\bf r}_2 =
{\bf r}'$ and  ${\bf r}_3 = {\bf r}_4 = {\bf r}''$, into the Eq.~(8)  and
integration over  the difference between ${\bf r}'$ and
${\bf r}''$. These coordinates are coupled by the interaction potential $V({\bf
r}' - {\bf r}'')$. Hence performing the integration we can exploit
Eq.~(\ref{K(e,p)}) only if the range of the interaction potential $d$ is
sufficiently short, $d\ll l_{\rm tr}, R_c$.

For a macroscopically homogeneous sample, classical probability
${\cal D}$ depends only on difference of its coordinates
${\bf r}= {\bf r}_2 -  {\bf  r}_1$. Using this fact and Eqs.~(\ref{deltaNu}) and
(\ref{K(e,p)}), we find:
\begin{eqnarray}\label{deltaNu10}
\frac{\delta \nu_{H} (\epsilon)}{\nu}
&=& -\frac{2}{\pi}  {\rm
Re}\! \! \int_{\epsilon}^\infty  \!\!\!\! d\omega \!\! \int \!\! \frac{
d\phi_1d\phi_2}{(2\pi)^2} \frac{\partial {\cal D}(\omega; {\bf
r}=0;\phi_1,\phi_2)}{\partial
\omega} \nonumber \\
&\times & V ( 2p_F | \sin [(\phi_1-\phi_2)/2]|),
\end{eqnarray}
where $V({q})$ is a Fourier transform of the interaction potential.
Formula (\ref{deltaNu10}) can be easily generalized for the case of the
long range Coulomb potential
$V({q})=2\pi e^2/\kappa q$ (here
$\kappa$ is the  dielectric constant). For such a potential, $V(q)$ in  Eq.
(\ref{deltaNu10})
should be replaced (see {\it e. g. } Ref. \cite{Altshuler}) by the screened
potential  $V_{\rm scr}( {\bf
q})=2\pi e^2/[\kappa (q + 2/a_B)]$ with  $a_B=\hbar^2\kappa/m e^2$ being the
effective Bohr radius.   Note, that the range of screened Coulomb potential,
$V_{\rm scr}(r)$, is of the order of the effective Bohr radius, and, therefore,
much smaller
than $l_{\rm tr}$ and $R_c$. Integration over the frequency in Eq. (\ref{deltaNu10})
immediately gives now the resulting expression for the Hartree contribution:
\begin{eqnarray}\label{deltaNu20}
 \frac{\delta \nu_{H} (\epsilon)}{\nu}
&=& \frac{2}{\pi}  {\rm
Re} \!\! \int \!\! \frac{
d\phi_1d\phi_2}{(2\pi)^2}{\cal D}(\epsilon; {\bf
r}=0;\phi_1,\phi_2)\nonumber \\
&\times & V_{\rm scr}( 2p_F| \sin[(\phi_1-\phi_2) /
2] | ),
\end{eqnarray}

The exchange correction to DOS can be obtained  in a similar  way as
 Eq. (\ref{deltaNu20}). The difference is, however, that in the exchange
counterpart of Eq. (\ref{deltaNu10}), one should put the retarded screened
potential $V_{\rm scr}(\omega, {\bf q})$ instead of $V(q)$.
The retardation of the interaction potential makes immediate integration over
frequencies, that led to  Eq.~(\ref{deltaNu20}),  no longer
possible. The resulting expression for the exchange correction has a form:
\begin{equation}\label{deltaNu21}
\frac{\delta \nu_{F} (\epsilon)}{\nu}
= \frac{1}{\pi}  {\rm
Re}\int_{\epsilon}^\infty  \!\!\!d\omega \!\! \int \!\! \frac{d{\bf q}}
{(2\pi)^2}V_{\rm scr}(\omega, {\bf q})
\frac{\partial \Delta(\omega; {\bf q})}{\partial \omega},
\end{equation}
where $V_{\rm scr}(\omega, {\bf q}) = V(q)/ [ 1+V(q) \Pi(q,\omega)]$, with
$\Pi(q,\omega) = \nu [1+i\omega \Delta (\omega ;{\bf q})]$  being the
polarization
operator, and
\begin{equation}\label{Sasha}
\Delta(\omega; {\bf q}) =  \int {d \phi_1\over 2\pi}\int{ d\phi_2\over
2\pi} d{\bf r} \, e^{-i{\bf qr}}\, {\cal D}(\omega; {\bf
r};\phi_1,\phi_2).
\end{equation}

Equations (\ref{deltaNu20}) and (\ref{deltaNu21}) express the interaction
correction to
tunneling DOS in terms of classical probability density ${\cal D}(\epsilon;
{\bf r};\phi_1,\phi_2)$ and are valid at  energies $\epsilon \lesssim E_F$.
Function ${\cal
D}(\epsilon; {\bf r};\phi_1,\phi_2)$ can be found from  the Boltzmann equation
describing the  scattering of electrons on impurities. In the special case
$\tau_s \ll
\tau_{\rm tr}$ we are interested in,  scattering on small angles dominates the
collision integral. With account for this simplification,  the transport
equation
takes the Fokker-Planck form:
\begin{eqnarray} \label{Boltzmann}
\Bigl(-i \omega & +&    {{\bf p}_2 \over m_e}  {\partial
\over
\partial {\bf r}} + \omega_c {\partial \over \partial  \phi_2} -
{1 \over \tau_{\rm tr}} {\partial^2  \over \partial \phi_2^2}\Bigr)
{\cal D}(\omega; {\bf r};\phi_1, \phi_2)  \nonumber \\
&=& 2\pi \delta(\phi_1-\phi_2) \, \delta ({\bf r}).
\end{eqnarray}
Equation (\ref{Boltzmann}) describes electron motion
along the cyclotron orbit accompanied by  angular diffusion caused by
scattering on a random potential. Analysis of Eq.~(\ref{Boltzmann}) yields:
\begin{equation} \label{sol1}
{\cal D} ({\omega}; {\bf q}; \phi_1, \phi_2)= \sum_n  {\cal D}_n ({\omega};
{\bf q}; \phi_1,
\phi_2),
\end{equation}
\begin{equation}
\label{sol2}
{\cal D}_n ({\omega}; {\bf q}; \phi_1, \phi_2) =  {e^{in(\phi_2-\phi_1)}
e^{i R_c {\bf
q} [({\bf p}_1 - {\bf p_2})/p_F\times {\bf z}]}
\over  -i(\omega- n\omega_c) + R_c^2q^2/2\tau_{\rm tr} + n^2 / \tau_{\rm tr}},
\end{equation}
where  ${\bf z}$ is a unit vector parallel to the magnetic field.  The solution
(\ref{sol1})-(\ref{sol2}) is valid for
$qR_c\ll\omega_c^2 \tau_{\rm tr}/(|\omega| + \omega_c)$.
The obtained solution of transport equation together with
Eqs.~ (\ref{deltaNu20}) and (\ref{deltaNu21}) enables  us to calculate the
interaction correction to DOS in the classical magnetic field.

At small frequencies, $\omega \ll \hbar/\tau_{\rm tr}$, the $n=0$  term in Eq.
(\ref{sol2}) contains a
part independent on initial, $\phi_2$,  and final, $\phi_1$,  directions of
the electron momentum.
This part dominates in Eq. (\ref{sol1}), so that ${\cal D} ({\omega}; {\bf
q}; \phi_1, \phi_2) \approx
1/(-i\omega + R_c^2q^2/2\tau_{\rm tr})$. This limit corresponds to the
diffusion regime studied in Refs.
\cite{Altshuler,Levitov}. Comparing Eqs. (\ref{deltaNu21}) and
(\ref{deltaNu20}), one
sees that in the diffusion regime the exchange correction to DOS contains
the interaction
potential, $V_{\rm scr}(\omega, k)$, at very small, $kR_c\lesssim 1$, momentum
transfers, while for the
Hartree correction the potential at large, $k\lesssim p_F$, momentum
transfers is important. Provided
that the interaction potential decreases rapidly with the increase of $k$, one
concludes
\cite{Altshuler} that in the diffusion regime,  $\epsilon \ll
\hbar/\tau_{\rm tr}$, the exchange
contribution to DOS, Eq.~(\ref{deltaNu21}),  dominates over the Hartree
one.  The resulting
expression for the interaction correction to DOS in the this regime yields:
\begin{equation} \label{ex1}
{\delta \nu(\epsilon) \over \nu} = -{\hbar (\omega_c \tau_{\rm tr})^2  \over
8\pi \epsilon_F \tau_{\rm tr} }
\ln\left({|\epsilon|\tau_{\rm tr}\over \hbar}\right)
\, \ln \left[{|\epsilon| \tau_{\rm tr}a_B^4 \over \hbar  R_c^4}\right].
\end{equation}
 Equation (\ref{ex1}) corresponds to the
Altshuler-Aronov-Lee
\cite{Altshuler} result with the renormalized by magnetic field diffusion
coefficient $D =
R_c^2/2\tau_{\rm tr}$.

At larger frequencies, $\omega \gg \hbar/\tau_{\rm tr}$, the probability ${\cal
D} ({\omega}; {\bf q};
\phi_1, \phi_2)$ describes quasiballistic motion of electron along the
cyclotron orbit. After the
period $2 \pi /\omega_c$ electron approaches  the vicinity of the initial
point with
a momentum only slightly deflected with respect to the initial direction.
According
to Eq. (\ref{deltaNu20}), that means that in the quasiballistic limit the
Hartree contribution to DOS
contain interaction potential at small momentum transfers. Thus, in
contrast to the diffusion regime,
there is no special reason why the Hartree contribution should be smaller
than the exchange one.
Furthermore, the Hartree correction appears to dominate over the exchange
one. Indeed, when comparing
the magnitudes of the Hartree and the exchange correction to DOS, one
should take into account
difference in a way how screening influences the two corrections. The
Hartree correction contains
screened interaction potential at zero frequency.   On the other hand,
expression for the exchange
correction, Eq. (\ref{deltaNu21}), contains integration over frequencies.
The polarization operator,
$\Pi(\omega, q)$, exhibits singularities at the same  frequencies as the
return probability, which
causes  suppression of the interaction potential $V_{\rm scr}(\omega, q)$, and,
in turn,
suppression of the  exchange contribution at energies close to multiples of
the cyclotron quantum.
As a result, it is the Hartree contribution\cite{foot} that determines the
peak structure of the
interaction correction to tunneling  DOS. Calculation  of this correction,
which consists in
substituting Eqs. (\ref{sol1})-(\ref{sol2}) and expression for $V_{\rm scr}$
into  Eq.
(\ref{deltaNu20}) and straightforward integration, gives  the resulting
expressions (\ref{result})
and (\ref{result1}). Peak structure  in tunneling DOS is  well-pronounced for
$n \lesssim \sqrt{\omega_c \tau_{\rm tr}}$. For larger $n$ width of the peaks
becomes comparable with $\hbar \omega_c$, and the oscillating structure of
DOS disappears.

Resulting energy dependence of the tunneling DOS of the interacting electron gas
in a classical magnetic field, obtained by numerical integration of Eqs.
(\ref{deltaNu20}) and (\ref{deltaNu21}) with account for Eq.~(\ref{sol1}), is
shown in Fig.~1. 

\begin{figure}
\centerline{\epsfxsize=3.2in\epsfbox{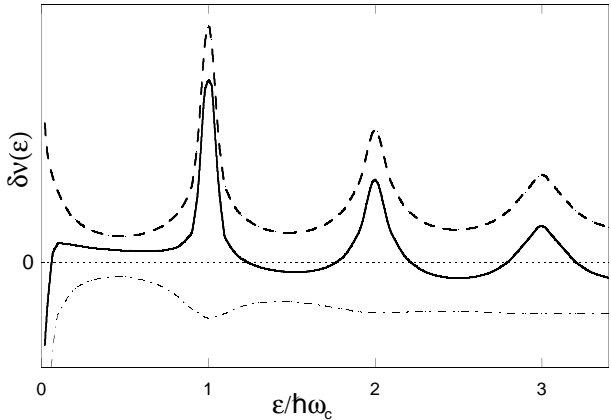}}
\narrowtext
\caption{Energy dependence of the exchange (dashed), the Hartree (dotted)
and the total (solid) interaction correction to tunneling DOS in a
classical magnetic field. Energy $\epsilon$ is measured from the Fermi
level. The curves are
calculated for $\omega_c\tau_{\rm tr} = 40$.}
\label{fig2}
\end{figure}

The range of magnetic fields $B$ where the oscillations of the
density of states $\nu (\epsilon)$ are caused by the interaction effects, is
confined by the condition $1/\tau_{\rm tr}\ll \omega_c \ll 1/\tau_{s}$, and
depends on the sample quality. In a high mobility sample with the typical
values\cite{Murphy} of relaxation times, $\tau_s\approx 10$ps and
$\tau_{\rm tr}\approx 300$ps this range is $10^{-3}{\rm T}\lesssim B \lesssim
5\cdot 10^{-2}{\rm T}$. Correspondingly, the typical numbers $N$ of observable peaks, $N\lesssim\sqrt{\tau_{\rm
tr}/\tau_{s}}$, for these samples is  $N\approx 6$.

To conclude, we have shown that classical ($\omega_c\tau_s \ll
1$) magnetic field  affects strongly the tunneling
density of states  of interacting electron gas. The tunneling DOS  is found
to be an
oscillating function of energy with the characteristic period  $\hbar
\omega_c$.

Discussions with  R.C. Ashoori and H.B. Chan are acknowledged with
gratitude. This work was
supported by NSF Grant DMR-9423244.

\end{multicols}


\begin{references}
\bibitem{foot} Calculation shows, that the magnitude of the exchange
correction near $\epsilon
\approx n\hbar \omega_c$ is  smaller than the Hartree term, Eq.
(\ref{result}), by a factor $\alpha_n
= [n^2/\omega_c\tau_{\rm tr}]^{1/3} \ll 1$.
\end{references}
\end{document}